\documentclass[sn-mathphys,Numbered]{sn-jnl}

\usepackage{graphicx}%
\usepackage{multirow}%
\usepackage{amsmath,amssymb,amsfonts}%
\usepackage{amsthm}%
\usepackage{mathrsfs}%
\usepackage[title]{appendix}%
\usepackage{xcolor}%
\usepackage{textcomp}%
\usepackage{manyfoot}%
\usepackage{booktabs}%
\usepackage{algorithm}%
\usepackage{algorithmicx}%
\usepackage{algpseudocode}%
\usepackage{listings}%
\usepackage{slashed}

\newcommand{\rmd}{\mathrm{d}}

\newcommand{\rmi}{\mathrm{i}}

\newcommand{\Tr}{\rm{Tr}}

\newcommand{\xp}{x^+} 			
\newcommand{\xm}{x^-} 			

\newcommand{\kp}{\boldsymbol{k}}
\newcommand{\qp}{\boldsymbol{q}}

\newcommand{\xv}{\boldsymbol{x}}
\newcommand{\rv}{\boldsymbol{x}}

\newcommand{\bv}{\boldsymbol{b}}

\newcommand{\qv}{\boldsymbol{q}}
\newcommand{\pv}{\boldsymbol{p}}

\newcommand{\xt}{x_T}			

\newcommand{\calO}{\mathcal{O}}
\newcommand{\calW}{\mathcal{W}}
\newcommand{\calP}{\mathcal{P}}

\newcommand{\zT}{\boldsymbol{0}_T}
\newcommand{\Dipper}{\textsc{McDipper}}

\definecolor{oscar}{RGB}{22, 156, 172}

\definecolor{oscarC}{RGB}{22, 156, 172}

\begin{document}
\raggedbottom
\title[Effective theories for nuclei
at high energies]{Effective theories for nuclei
at high energies}


\author*[1]{\fnm{Oscar} \sur{Garcia-Montero}}\email{garcia@physik.uni-bielefeld.de}
\equalcont{These authors contributed equally to this work.}

\author[1]{\fnm{Sören} \sur{Schlichting}}\email{sschlichting@physik.uni-bielefeld.de}

\equalcont{These authors contributed equally to this work.}

\affil[1]{\orgdiv{Fakultät für Physik}, \orgname{Universität Bielefeld}, \orgaddress{\street{Universitätstraße 25}, \city{Bielefeld}, \postcode{D-33615}, \country{Germany}}}

\abstract{We discuss the application of the Color Glass Condensate (CGC), an effective field theory of Quantum Chromodynamics (QCD), to describe high-energy nuclear interactions. We first provide an introduction to the methods and language of the CGC, its role in understanding gluon saturation in heavy-ion collisions at the LHC and RHIC, and its relevance in various scattering processes such as Deep Inelastic Scattering (DIS). The application of the CGC effective field theory to describe hadron-hadron collisions is discussed in the scope of asymmetric \textit{dilute-dense} collisions, and Heavy-Ion Collisions in the \textit{dense-dense} limit. The review covers theoretical foundations, recent advancements, and phenomenological applications, focusing on using the CGC to determine the initial conditions of heavy-ion collisions.}

\keywords{Color-Glass Condensate, Saturation, Deep-Inelastic Scattering, Heavy-Ion Collisions, Nuclear Structure, Forward physics}



\maketitle

\section{Introduction}\label{sec:intro}


High-energy heavy-ion collisions at the Large Hadron Collider (LHC), and the Relativistic Heavy Ion Collider (RHIC) offer a unique window into the properties of nuclei at high energies and the behavior of nuclear matter under extreme conditions. Clearly, at such high energies, low-energy descriptions of atomic nuclei based solely on nucleons and mesons become inapplicable, giving way to other effective theories of Quantum Chromodynamics (QCD) that capture the emergent phenomena and interactions of quarks and gluons within the nuclear medium. Nevertheless, at the scales at which nuclear structure manifests itself,  the theory of strong interactions remains strongly coupled and highly non-perturbative. Hence, to describe the relevant properties of nuclei at high-energies, effective theories based on the principles and symmetries of QCD have been developed. Such is the case of the Color Glass Condensate~\cite{Gelis:2010nm,McLerran:1993ni,McLerran:1993ka,McLerran:1998nk}, an effective 
theory of QCD in high-energy scattering processes. 

Based on the fundamental degrees of freedom of QCD, the CGC accounts for the rich interplay between linear production and non-linear absorption of gluons, which marks the onset of collective gluon dynamics in the boosted nucleus and describes the phenomenon of gluon saturation as an emergent property. The scope of this review is to provide a pedagogical introduction to the CGC effective theory, and its applications to Deeply Inelastic electron-proton $e+p$ and electron-nucleus $e+A$ Scattering (DIS) and hadronic proton-proton $p+p$, proton-nucleus $p+A$ and nucleus-nucleus $A+A$ collisions. Below, in the remainder of this section, we lay out the basic concepts of high-energy scattering in which the CGC effective theory is framed.

Subsequently, in Sec.~\ref{sec:CGC}, we sketch the basic theoretical ideas underlying the CGC, and discuss the application to DIS in Sec.~\ref{sec:DIS}. We proceed by summarizing the advances and status quo of the CGC in the context of hadronic collisions in Sec.~\ref{sec:HICs}, where in Sec.~\ref{sec:forward} we focus on the so-called dilute dense limit, e.g. used in the case of forward particle production in p+A collisions, while in Sec.~\ref{sec:dense} we describe the dense-dense limit, usually employed to calculate the initial conditions for the space-time of the medium created in heavy-ion collisions. We conclude this review in Sec.~\ref{sec:conclusions}.


\subsection{High energy scattering in QCD}
\label{subsec:HighEnergyScatQCD}
The fundamental subnucleonic degrees of freedom of hadrons, quarks and gluons, create a rich substructure inside nuclei, manifested by a complex many-body nucleonic (nuclear) QCD wavefunction. High-energy scattering experiments such as DIS are used to access this partonic QCD content. 
The information of the nuclear wavefunction can be understood in terms of correlation functions of (anti-)quark $\Psi$ fields and gluon field-strength operators, $F^{\mu\nu}_{a}$. Such QCD correlation functions are in general very complex, as they reflect all the information about the configurations of colored particles inside hadrons or nuclei. Nevertheless, there are well established theoretical and experimental formulations for two-point correlations functions, such as collinear parton distribution functions (PDFs), Transverse Momentum Distributions (TMDs), Generalized Parton Distribution Functions (GPDs), Wigner Distributions, and we refer to \cite{Ethier:2020way,Diehl:2015uka,Ji:2004gf,Boussarie:2023izj} for dedicated reviews on the subject. The simplest case is 
the collinear gluon PDF $f_{g/h}(x,\mu^2)$ inside a hadron/nucleus $h$, given by
\begin{equation}
    f_{g/h}(x,\mu^2) = \int \frac{\rmd y^-}{2\pi}\frac{e^{-ixP^+_h y^-}}{x P^+_h}\left\langle P^+_h,\zT|F_a(0,y^-,\zT)^{\nu+}\calO^{ab}\,F_b(0,0,\zT)_{\nu}^{+}|P^+_h,\zT\right\rangle_{\overline{MS}}\,,
\end{equation}
which describes the  distribution  of gluons with (light-cone) momentum fraction $x=p^{+}_g/P^{+}_h$, measured at a renormalization scale $\mu^2$. The integration is performed along the light-cone position $y^-$, conjugate to the light-cone momentum $P^+$. In what follows, we will be using the light cone positions and momenta, which are defined 
by $x^{\pm} =(t\pm z)/\sqrt{2}$ and $p^{\pm} =(p_0\pm p_z)/\sqrt{2}$, respectively.
In a collinear PDF there is no information on the intrinsic transverse momentum of the gluons, since the field-strength correlator is integrated over transverse momenta, i.e. the gluonic field strength is measured at the same transverse position $\zT$. This is different for a TMD, where this information is retained~\cite{Boussarie:2023izj}. Similarly GPDs retain information about the spatial position of partons inside hadrons~\cite{Diehl:2003ny}, and Wigner Distributions retain information about position and momentum~\cite{Ji:2004gf,Belitsky:2003nz}.  Irrespective of this, it is important to note that all PDFs,TMDs, GPDs,  only contain information on single parton distributions and do not encapsulate the full extent of the many-body complexity of the nuclear wavefunction. 

Due to the non-perturbative nature of QCD bound states, QCD correlation functions such as TMDs and PDFs, cannot be computed from first principles in perturbative QCD. Such non-perturbative knowledge  
has been historically extracted from from fits to observables measured in experiments such as DIS at the HERA collider at DESY, Hamburg and the Thomas Jefferson National Accelerator Facility (JLAB).  More modern extractions perform a simultaneous extraction including also data from hadronic collisions such as jet, high-$k_T$ charmed meson, and electromagnetic probe data~\cite{Klasen:2023uqj}. Theoretical advances and algorithmic improvements in lattice QCD, can also help to provide additional theoretical insight into the behavior of these non-perturbative QCD correlation functions. For a recent review on this rapidly evolving field, we refer to Ref.~\cite{Cichy:2018mum}.


\subsubsection{Kinematics}
\label{sec:kinematics}

To gain a more intuitive understanding of the partonic content of hadrons and nuclei, it is insightful to consider a high-energy scattering experiment, such as DIS, where a probe interacts with the partons inside hadrons, exchanging four-momentum $q$ in the process. Generally speaking, the kinematics of this process can be described in terms of the center-of-mass (CoM) energy $\sqrt{s}$, the virtuality of the photon, $Q^2=-q^{2}$, and the Bjorken scaling variable $x$. 
Intuitively, $Q$ provides a \textit{resolution scale} for the partonic content of the hadron, while one can relate $x$, in single parton kinematics, to the momentum fraction carried by the interacting parton. Additionally, $x$ is related to the inelasticity of the process $y=Q^2/sx$.

Different regions in the $(x,Q^2)$ phase-space, can then be accessed depending on the detailed kinematics of the scattering experiment. 
For a fixed center-of-mass energy, more massive final states (e.g. $Z$-bosons) sample regions of larger $x$, whereas higher center-of-mass energies give access to smaller values of $x$. Generally, one finds that at typical values of $Q^2 \sim {\rm GeV}$, hadrons consist of a small number of large $x\sim 1$ valence partons, and a large number of sea partons with momentum fractions $x \ll 1$.

Even though the property of asymptotic freedom guarantees that the underlying QCD interactions can be treated perturbatively at high momentum transfer, the complex partonic structure of hadrons and nuclei and the non-linear nature of gluon interactions make a general description of high-energy scattering processes intractable. Depending on the kinematic regime that is being probed, different theoretical formalisms/expansion schemes in QCD can then be used to describe the process. The traditional and most well established description of perturbative QCD, known as collinear factorization,  describes scattering events comprised of small numbers of hard scatterings of partons by approximating the hadronic cross sections in terms of a convolution of partonic cross sections with the non-perturbatively extracted collinear PDFs. The separation of non-perturbative dynamics inside the hadron and the perturbative dynamics of partonic scattering, is known in the literature as the \textit{factorization theorem}, which exploits a separation of scales that also allows one to express the scattering processes in terms of scatterings of a few partons.  While the separation of nucleonic and partonic scales guarantees a good theoretical control, it also means that there is a  priori no connection to low energy nuclear structure, e.g. the spatial distribution of nucleons in nuclei. Evidently, the pQCD approach requires small values of 
the coupling constant of QCD, $\alpha_S(Q^2)\ll 1$; however due to the limitation to a small number of partonic scatterings the range of applicability of this approach is also limited to large/moderate values of $x$, where hadronic structure is dominated by few partons. 

\subsection{Gluon saturation and the Color Glass Condensate}\label{subsec:CGC}
Conversely, when $x\ll1$ (i.e. low-$x$) the collinear parton distributions of gluons and sea-quarks increase according to a power law $\sim x^{-a(Q^2)}$, due to the radiative emissions of soft gluons, carrying a small fraction of the hadron's momentum. Naturally, the conventional perturbative description breaks down when gluon densities become non-perturbatively large $\sim 1/{\alpha_S}$, and new effective descriptions of QCD are needed to account for the multi-particle dynamics inside the hadron. While in principle there are different possibilities to account for the ensuing non-linear QCD dynamics, we will focus on the Color Glass Condensate effective field theory of high-energy QCD, which provides a weak coupling description of high-energy scattering processes, in the multi-parton regime at low-$x$.

Clearly, one of the most remarkable features of the Color Glass Condensate Effective Field Theory is that it predicts the feature of perturbative gluon saturation in hadrons and nuclei. While in the few parton regime, the evolution of parton densities towards smaller values of $x$ can be calculated perturbatively and gives rise to the linear BFKL evolution~\cite{Kuraev:1976ge,Kuraev:1977fs,Balitsky:1978ic}, the phase-space density of gluons increases with decreasing $x$ due to successive radiative emissions. Consequently, at small-$x$, gluons densely populate the available phase-space, and it becomes increasingly likely for gluons to recombine -or fuse-, leading to a non-linear evolution of the parton densities, as encoded in the BK/JIMWLK evolution equations of small-$x$ QCD~\cite{Balitsky:1995ub,Kovchegov:1999yj,Iancu:2001ad,Iancu:2000hn,Iancu:2001md,Mueller:2001uk,Rummukainen:2003ns}. Eventually, the balance between radiative gluon emission and recombination leads to a saturation of the gluon density in the multi-parton regime at small-$x$.


More precisely, the interplay between radiative emissions and absorptions creates an effective semi-hard scale, $Q_s(x)$, which serves as a dynamical scale separating the two regimes. Specificallxy, for transverse momenta, $k_T<Q_s$, the gluon distribution is saturated while gluons with higher transverse momenta, $k_T>Q_s$ are still dilute in phase space. Clearly, the saturation scale $Q_{s}(x)$ is a dynamical scale as it 
is controlled by the variable $x$, where phenomenologically it has been found that $Q_s^2(x)\sim x^{-\lambda}$, with $\lambda\approx 0.2-0.3$~\cite{Golec-Biernat:1998zce}. Since the nucleon density affects the saturation scale, larger nuclei become saturated already at lower energies, and the saturation scale $Q_{s,A}$ in nuclei is enhanced compared to the $Q_{s,Ap}$ saturation scale in a proton, as $Q_{s,A}^2\sim A^{1/3}Q_{s,p}^2$~\cite{McLerran:1993ni,McLerran:1993ka}. Hence, the Color Glass Condensate framework is particularly well suited for an effective description of nuclei at high-energies, where a larger fraction of gluons have transverse momenta $k_T \lesssim Q_s$ and we will discuss its phenomenological applications in Sec.~\ref{sec:DIS} and~\ref{sec:HICs}.




\section{Effective theory for high-energy QCD -- Color Glass Condensate}\label{sec:CGC}
\textbf{Basics:} Despite the fact that due to their complicated partonic structure the scattering of QCD bound states represents a highly complicated process, even at asymptotically high energies, it is instructive to follow Bjorken~\cite{Bjorken:1968dy}  and first consider the much simpler process of high-energy scattering of partons. We picture a process where an incoming quark with large momentum interacts with the color field of the target. 
\begin{wrapfigure}{r}{0.4\textwidth}
\includegraphics[width=0.4\textwidth]{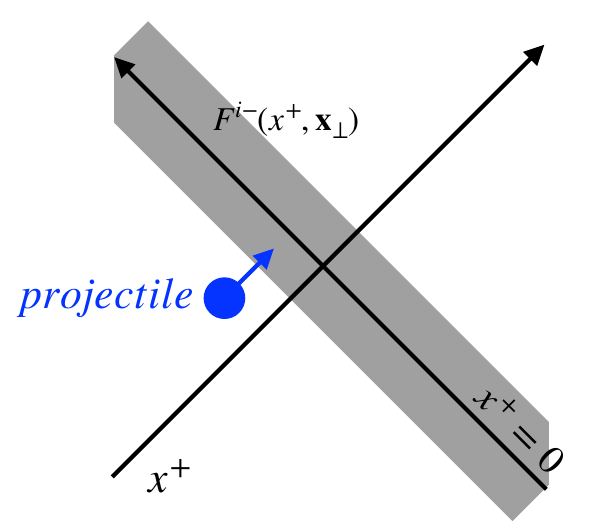}
\caption{Space-time picture of high-energy scattering\label{fig:ScatteringCartoon}}
\end{wrapfigure}
In the laboratory frame we can set the incoming quark to be moving to the right and the target to be moving to the left, both at almost the speed of light, as illustrated in Fig.~\ref{fig:ScatteringCartoon}. Due to Lorentz contraction,  the influence of the color fields of the target is then confined to a narrow region around $\xp=\frac{t+z}{\sqrt{2}}=0$. Similarly, due to time dilation, the color fields of the target appear as quasi-static i.e. approximately $\xm=\frac{t-z}{\sqrt{2}}$ independent fields, such that if we adapt an appropriate gauge choice and view the color field as emerging from individual color charges, with density $\rho^{a}(\xp,\xt)$, the color field of the target can be expressed as~\cite{McLerran:2001sr}
\begin{align}
    A^{-}(\xp,\xt)=\frac{1}{-\Delta_\perp} g\rho^{a}(\xp,\xt)t^{a}
\label{eq:single_nucleus_cgc}
\end{align}
with field-strength $F^{i-}(\xp,\xt)=\partial^{i}A^{-}(\xp,\xt)$ localized around $\xp\sim 0$, where $\xp$ can be viewed as a time variable in the process. The operator $\Delta_\bot$ corresponds to the inverse transverse gluon propagator, and therefore, acting with its inverse is equivalent with a convolution with the transverse gluon propagator. Since the interaction of projectile and target only takes place around $\xp\simeq 0$, the incoming/outgoing quark propagates in the vacuum before $(\xp<0)$ and after $(\xp>0)$ the interaction with the color field of the target, and the non-trivial part of the calculation is to relate the fermion field  $\Psi(\xp=0^{+},\xm,\xv_T)$ after the interaction $(\xp=0^{+})$ to the incoming field $\Psi(\xp=0^{-},\xm,\xv_T)$. By treating the problem to lowest order in perturbation theory, the interaction of the quark with the target is then described by the solution of the Dirac equation $(i\slashed{D}-m)\Psi(x)=0$, which to leading order in the eikonal approximation takes the form
\begin{align}
(\partial_{+} -ig A^{-}(x)) \Psi^{-}(x)=0\;,
\label{eq:Dirac}
\end{align}
for the relevant light cone component $\Psi^{-}=\frac{1}{2}\gamma^{-}\gamma^{+}\Psi$ of the Dirac field (see e.g. \cite{McLerran:1998nk} for details). The solution of Eq.~\eqref{eq:Dirac}, 
\begin{align}
    \Psi^{-}(\xp=0^{+},\xm,\xv_T)=V(\xv_T)  \Psi^{-}(\xp=0^{-},\xm,\xv_T)
    \label{eq:Dirac2}
\end{align}
shows that high-energy scattering processes in QCD are described by light-like Wilson lines, 
\begin{equation}
    V(\xv_T)=\calP \exp\left[ig \int_{-\infty}^{\infty}\rmd z^+\, A^{-}(z^{+},\xv_T) \right]\,, 
    \label{eq:wilson_line}
\end{equation}
which characterize the eikonal propagation of color charges. By performing a Fourier analysis of the incoming and outgoing quark field, we immediately see that -- in addition to a color rotation -- the 
light-like Wilson line $V(\xt)$ characterizes the transverse momentum transfer to the quark, as 
\begin{align}
    \Psi^{-}(\xp=0^{+},\xm,\pv_T)=\int \frac{d^2\qv_T}{(2\pi)^2} \tilde{V}(\qv_T) \Psi^{-}(\xp=0^{-},\xm,\pv_T-\qv_T)
    \label{eq:Dirac3}
\end{align}
such that the momentum $\pv_T$ of the outgoing quark is given by the sum of the incoming momentum $\pv_T-\qv_T$ and the momentum transfer $\qv_T$ from the target.
Since the Wilson-line $V(\xv_T)$ contains an infinite number of gauge field insertions, it re-sums the transverse momentum transfer encountered in multiple interactions with the target, while neglecting the longitudinal momentum transfer in all interactions~\cite{Dumitru:2018kuw}. Conversely, in the dilute limit, where the Wilson line can be expanded around $V(\xv_T)=\mathbb{I}$, one can recover the high-energy limit of the standard perturbative QCD result\footnote{Note that in the high-energy limit, the longitudinal momenta are large and the small longitudinal momentum transfer that is included in the pQCD calculation can be neglected. The calculation sketched above, is directly performed in the high-energy limit and neglects longitudinal momentum transfer from the very beginning.} (see Refs.~\cite{Petreska:2018cbf,Mulders:2000sh,Marquet:2016cgx,Kotko:2015ura, Garcia-Montero:2023gex} and references therein).

Despite the simplistic nature of the above example, it serves as a good illustration of the underlying physics. Hence, the general strategy of perturbative QCD (pQCD) calculations is to separate the dynamics of the partonic sub-process  from genuinely non-perturbative information such as the distribution of incoming quarks in a hadronic bound state. In this case, such partonic dynamics would be the high-energy scattering of the quark with the color field of the target. By supplementing the required non-perturbative information, about a) the (momentum-) distribution of partons inside the projectile in form of a collinear PDF and
b) the correlation functions of light-like Wilson lines inside the target, the above expressions can thus be turned into a leading order cross-section calculation for forward particle production, as discussed in Sec.~\ref{sec:forward}. 

In this spirit, the Color-Glass Condensate (CGC) is an effective theory for high-energy scattering
based on light-like Wilson lines as fundamental degrees of freedom. Generically, the computational strategy is to compute observables, such as production cross-sections or particle production yields, in the presence of an individual realization of the Wilson lines $V(\xv)$ of a dense hadronic object. Subsequently, one performs a statistical average over all possible realizations of the Wilson lines $V(\xv)$, as described by a weight functional $W_{x}[V]$
\begin{equation}
    \left\langle \calO\right\rangle_x= \int DV \, \calW_x[V]\,\calO[V]
    \label{eq:evolution}
\end{equation}
Generally speaking, the weight functional $W_{x}[V]$ encodes information about correlated scattering of multiple highly energetic partons, and is not known a priori. Moreover, since this theoretical description requires a hitherto arbitrary separation into gluon fields that generate the Wilson lines, and dynamical gluons which enter perturbative calculation of matrix elements, there is a requirement that cross-sections are invariant under a change of this separation scale. This process gives rise to the JIMWLK evolution equation~\cite{Jalilian-Marian:1996mkd,Jalilian-Marian:1997qno,Jalilian-Marian:1997jhx,Iancu:2000hn,Iancu:2001md}, which describes the evolution of the weight functional $W_{x}[V]$ with decreasing momentum fraction $x$ or increasing center-of-mass energy for fixed kinematics. 

By now the most frequently adopted strategy is thus to develop a physically motivated model for the weight functional at an initial scale, typically $x_0 \sim 0.1$, calculate its evolution with decreasing $x$ based on JIMWLK (or BK) evolution, and constrain the parameters of the model by fits to deep inelastic scattering or hadronic collision data. By asserting universality of the CGC weight function, the same theoretical framework and model can then be used to provide a unified description of different high-energy
scattering experiments from deep-inelastic scattering (DIS) to complex hadronic collisions.\\

\begin{wrapfigure}{r}{0.4\textwidth}
\includegraphics[width=0.35\textwidth]{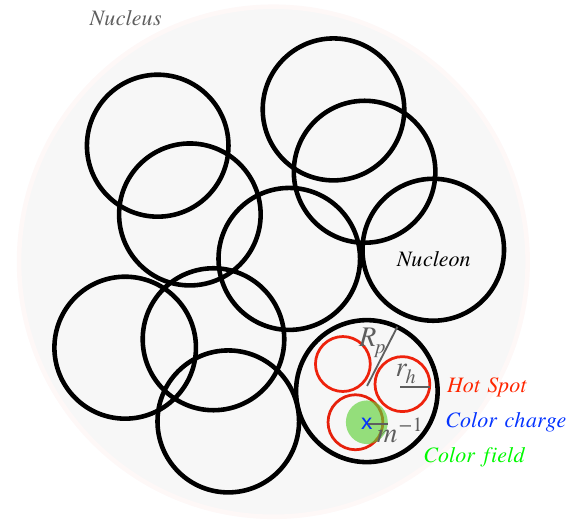}
\caption{Illustration of the color fields of a nucleus in an impact-parameter dependent model of the target. }
\label{fig:IPSatCartoon}
\end{wrapfigure}
\noindent\textbf{IP-Glasma model:} A common example is the phenomenologically  successful IP-Glasma model~\cite{Schenke:2012wb,Schenke:2012hg,Schenke:2016ksl,Schenke:2018fci}, for which the basic anatomy of the model is depicted in Fig.~\ref{fig:IPSatCartoon}. Starting from the large scale structure, nucleons are statistically distributed inside the nucleus according to nuclear structure input, which requires the simultaneous knowledge of the fluctuating position of all nucleons in a given realization of the nucleus\footnote{Note that the nucleon positions are assumed to remain constant over the time scale of a high-energy scattering event.}. Each nucleon then contains a fluctuating color charge density $\rho$, concentrated around $N_q$ hot spots with a (transverse) spatial size $r_H$, which are statistically distributed within the size of the nucleon $R_p$. By virtue of Eq.~\eqref{eq:single_nucleus_cgc}, the color charges $\rho$ create color fields $F^{i-}$, whose large distance behavior is regulated by an infrared regulator $m \sim \Lambda_{\rm QCD}$, such that the color field of an individual charge extends over a transverse distance $\sim m^{-1}$. It is important to point out that the model also provides information on the transverse spatial structure of the field strength beyond the overall magnitude of field strength fluctuations as a function of $x$, which is traditionally encoded in collinear parton distribution functions. This new information can be related to generalized parton distributions (GPDs) and transverse correlations of the field strength relevant to transverse momentum dependent (TMD) parton distributions~\cite{Diehl:2003ny,Dominguez:2011wm,Marquet:2016cgx}.

The IP-Glasma model provides the magnitude of color charge fluctuations $\rho(\xv)$ at an initial scale $x_0 \sim 0.1$, such that the subsequent evolution towards smaller values of $x < x_0$ is described by the JIMWLK evolution equation~\cite{Mantysaari:2016ykx}\footnote{A variant of this is the succesful IP-Sat model, where $\rho(\xv,x)$ is provided directly as a function of $x$~\cite{Kowalski:2006hc,Rezaeian:2012ji}. More specifically, $\log\Tr[V(\bv+\mathbf{r}/2)V^\dagger(\bv-\mathbf{r}/2)]= 
\frac{\pi^2}{2N_c}r^2\alpha_S(\mu^2)xg(x,\mu^2)T(\bv) $, where the scale $\mu^2=\mu^2_0+C/r^2$. As the reader can see, the correlation function contains local information, probed at the average position, $|\bv|$, while the relative direction, $|\mathbf{r}|$ sets the scale at which the target is probed.}. Since the model features various kinds of fluctuations, from fluctuating color charge densities at the smallest scales to nucleon positions at the largest scales, analytic calculations are usually only possible e.g. in the dilute limit~\cite{Demirci:2021kya,Demirci:2022wuy}, and expectation values of observables are typically calculated by statistically sampling the averages in Eq.~\eqref{eq:evolution} based on a Monte Carlo procedure~\cite{Mantysaari:2018zdd}. Each realization then provides a particular configuration of the Wilson lines $V_{x}(\xv)$, such that also $n$-point correlation functions of Wilson lines can easily be calculated~\cite{Marquet:2016cgx,Lappi:2017skr,Deganutti:2023qct}, and in many cases individual realization are interpreted as corresponding to single events (c.f. Sec.~\ref{sec:dense}).

\begin{figure}[t!]
    \centering
\includegraphics[scale=0.18]{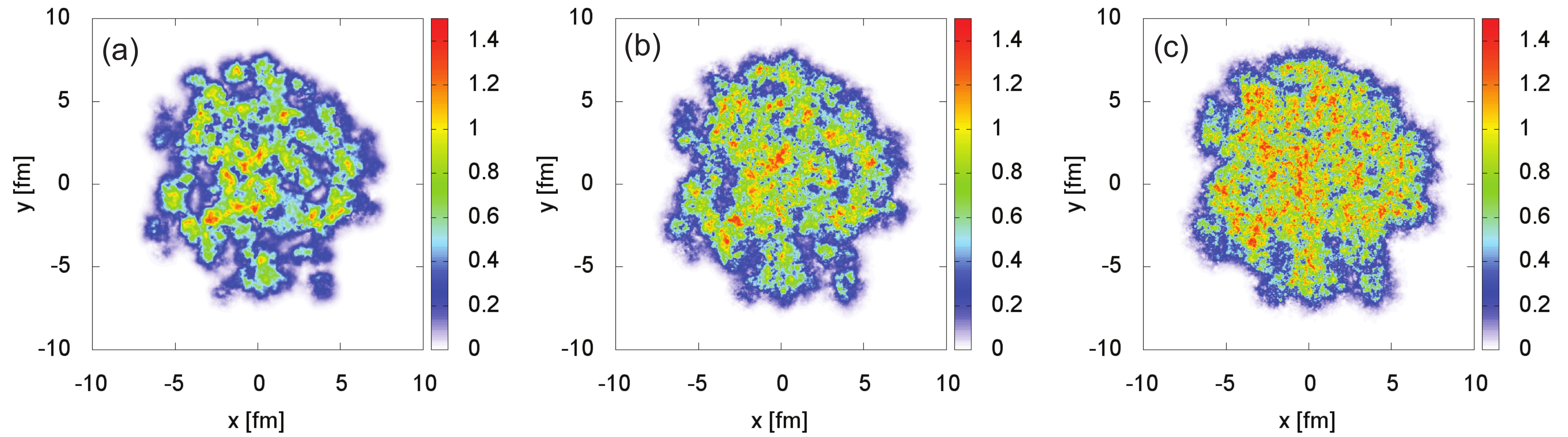}
    \caption{Color fields of a $Pb^{208}$ nucleus in the IP-Sat model at rapidities  $x_0\approx 2\times 10^{-3}$, with $Y_0=-2.4$ (left) and after $\Delta Y=2.4$ (middle) and $\Delta Y=4.8$ (right) units of small-$x$ JIMWLK evolution. Figure taken from Ref.~\cite{Schenke:2016ksl}}
    \label{fig:JIMWLKNucleus}
\end{figure}

We illustrate this procedure in Fig.~\ref{fig:JIMWLKNucleus}, which depicts a realization of the color fields of a $Pb^{208}$ nucleus, as obtained a) at the initial scale $x_0$ and (b),(c) after $\Delta Y=\log(x_0/x)=2.4,4.8$ units of (fixed coupling) JIMWLK evolution~\cite{Schenke:2016ksl}. The evolution towards smaller $x$ is characterized by an increase of the saturation scale which is calculable perturbatively~\cite{Triantafyllopoulos:2002nz} and reflected by the smaller and smaller scale structures visible in Fig.~\ref{fig:JIMWLKNucleus}. Beyond that the small-$x$ evolution also inevitably leads to a growth of the transverse spatial extent of hadrons and nuclei, although the incorporation of such impact parameter dependent effects necessarily requires some additional modeling to regulate large distance emission~\cite{Kovner:2003zj,Mantysaari:2022ffw,Schenke:2015aqa}. In particular, this process of Gribov diffusion~\cite{Gribov:1983ivg}, also leads to a change of the large-scale structure of the color fields which can be seen by comparing the left and right panels of Fig.~\ref{fig:JIMWLKNucleus}. While initially the color fields of the nucleus are concentrated inside the nucleons, the nucleus becomes increasingly opaque and as structure of individual nucleons is washed out towards smaller $x$. It would be interesting to explore, whether evidence for the predicted change of the large scale nuclear structure can indeed be observed in very forward or very high energy scattering experiments, that probe sufficiently small values of $x$.

\section{Probing the nuclear wavefunction: DIS}
\label{sec:DIS}
\begin{figure}[t!]
\includegraphics[width=0.45\textwidth]{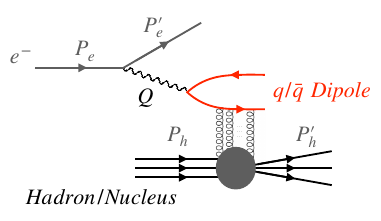}
\hspace{1cm}
\includegraphics[width=0.33\textwidth]{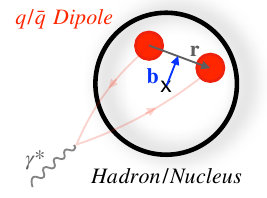}
\caption{Deep-Inelastic scattering (DIS) in the dipole picture. \label{fig:DISCartoon}}
\end{figure}

Deep-inelastic $e+p/A$ scattering experiments provide a theoretically clean way to probe the partonic structure of nucleons and nuclei. In the high-energy limit, this process is best viewed in the dipole picture~\cite{Kovchegov:2012mbw} where, as illustrated in Fig.~\ref{fig:DISCartoon}, the electron emits a virtual photon, which subsequently splits into a quark/anti-quark pair that interacts as a color dipole with the hadronic target. Due to this separation of time scales between QED and QCD processes, the leading order cross-section for the inclusive scattering of a  virtual photon $\gamma^{*}$ is given by 
\begin{equation}
    \sigma_{L,T}^{\gamma^*p}(Q^2,x)=2\sum_f \int \int \rmd^2 \bv \,\rmd^2 \rv \int \rmd z \left|\Psi_{L,T}^{(f)}(r,z;Q^2)\right|^2 D(x,b,r)
\end{equation}

The cross section is simply determined by a convolution of 
the QED part $\left|\Psi_{L,T}^{(f)}(r,z;Q^2)\right|^2$, representing the probability for the longitudinally (L) or transversely (T) polarized virtual photon with virtuality $Q^2$ to split into a $q/\bar{q}$ dipole with momentum fractions $z,1-z$ and transverse separation $r$. The remaining ingredient is the color dipole cross-section
\begin{equation}
D(x,b,r)=\left\langle\frac{1}{N_c}{\rm tr}\left[1-V(b+r/2)V^\dag(b-r/2)\right] \right\rangle_{x}
\end{equation}
representing the interaction of the quark/anti-quark dipole of size $r$ at an impact parameter $b$ with the hadronic target~\cite{Kovchegov:2012mbw}. 

Statistical comparisons to measurements of the inclusive $\gamma^{*}p$ (reduced) cross-section at the HERA collider~\cite{H1:2009pze} as a function of $x$ and $Q^2$, then allow one to constrain the $x$ and $r$ dependence of the dipole scattering amplitude, as pioneered early on in~\cite{Golec-Biernat:1999qor,Kowalski:2003hm,Kowalski:2006hc}. However, the comparison to inclusive cross-sections suffers from the deficiency, that even at moderate values of $Q^2 \sim {\rm GeV}$ it is sensitive to the contributions of non-perturbatively large dipoles ($r \gtrsim \Lambda_{\rm QCD}^{-1}$) and does not provide information on the (transverse) geometry, associated with the impact parameter $(b)$ dependence of the dipole amplitude. Hence it is advantageous to consider more exclusive processess, such as e.g. the exclusive production of (heavy) vector mesons in $\gamma^*+p \rightarrow J/\psi+p$ reactions\footnote{Similarly one could also consider the  $e+p \to e+p +\gamma$ deeply virtual Compton scattering (DVCS) or $e+p \to e+p +J+J$ diffractive di-jet production processes, and we refer the interested reader to~\cite{Salazar:2019ncp} (and references therein) where these processes are discussed.}, where the momentum transfer $t$ encodes information on the impact parameter $(b)$ dependence of the dipole amplitude~\cite{Kowalski:2006hc}. 
Exclusive vector meson production in DIS experiments has been studied at HERA~\cite{H1:1999pji,ZEUS:2005bhf,ZEUS:2004yeh,H1:2005dtp}, and one distinguishes between coherent and incoherent processes, where the proton remains intact or disintegrates, which according to the Good-Walker picture~\cite{Good:1960ba}, probe the average and fluctuations of the interaction. Experimental measurements of \emph{coherent} vector meson production at small momentum transfer $|t| \lesssim 1{\rm GeV}^2$ put a tight constraint on the average spatial extent $R$ of the nucleon~\cite{Mantysaari:2016ykx,Mantysaari:2017dwh,Demirci:2022wuy,Golec-Biernat:1999qor}\footnote{Since $R$ is extracted from a process at low-$x$, using colored probes, this radius is to be thought of as a gluonic radius. This is in juxtaposition, for example to the charge radius of the proton as extracted from elastic scattering experiments~\cite{Gao:2021sml} }, and also show intriguing hints of the energy dependence of the proton size due to small-$x$ evolution~\cite{Mantysaari:2016ykx,Mantysaari:2016jaz,Schlichting:2014ipa}. Conversely, at larger momentum transfers $|t|\gtrsim 1 {\rm GeV}^2$, the cross-section is dominated by the \emph{incoherent} process, the experiments also clearly demonstrate the importance of fluctuations at the subnucleonic scale, which in phenomenologically successful models is incorporated through sub-nucleonic hotspots and color charge fluctuations~\cite{Demirci:2021kya,Demirci:2022wuy}, as illustrated in Fig.~\ref{fig:IPSatCartoon}.

Similarly, to such imaging of the proton, it is also possible to probe nuclear structure at high energies, by considering deep-inelastic scattering processes off nuclear targets. While so far there is only a rather limited amount of experimental data for nuclear targets, this will change with the operation of the Electron Ion Collider (EIC),  and recent works have started to develop the phenomenology of the effects of nuclear deformation and correlations of nucleons~in nuclear DIS \cite{Mantysaari:2023qsq,Mantysaari:2023prg,Mantysaari:2023gop}. Beyond deep-inelastic scattering, which provides a theoretically clean probe of nuclear structure at high energies, such imaging of protons and nuclei by exclusive vector meson production can also be explored in so-called ultra-peripheral collisions (UPCs) of protons and nuclei (p+p,p+A,A+A) at RHIC and LHC, where the hadronic interaction is mediated by photon exchange~\cite{Baltz:2007kq,Klein:2017vua}. While the calibration of the photon flux, and the quantum interference between the photoproduction off either of the two colliding hadrons complicate the theoretical interpretation of UPCs, sophisticated techniques have recently been developed to extract $\gamma + A \to J/\Psi +A$ cross-sections from UPC measurements~\cite{ALICE:2022iqi,Klein:2020fmr,Guzey:2013jaa}. Naturally, these UPC measurements~\cite{ALICE:2013wjo,ALICE:2021tyx,ALICE:2023jgu,CMS:2019awk,CMS:2023snh,LHCb:2022ahs,STAR:2023nos} provide access to different $\gamma+A$ center of mass energies ($W$), which can provide additional information on the energy dependence of nuclear and nucleon structure and the magnitude of gluon saturation effects. 
\subsection{New developments \& open questions}
New developments in the context of DIS primarily concern the quests for increasing precision and new phenomenological applications at the upcoming Electron Ion Collider (EIC)~\cite{Accardi:2012qut}. A substantial amount of effort has been dedicated to improve calculations of $e+h$ impact factors to NLO precision for observables such as DIS structure functions~\cite{Beuf:2020dxl,Beuf:2021qqa,Balitsky:2012bs,Beuf:2011xd,Beuf:2016wdz,Beuf:2017bpd,Lappi:2016oup,Hanninen:2017ddy,Ducloue:2017ftk}, exclusive di-jet production \cite{Boussarie:2016bkq,Boussarie:2016ogo,Boussarie:2019ero,Caucal:2023nci} and exclusive vector 
meson~\cite{Mantysaari:2021ryb,Mantysaari:2022bsp}. We list also recent advances in the physics of dijets/dihadron photoproduction, \cite{Caucal:2021ent,Caucal:2023fsf,Taels:2022tza,Bergabo:2022tcu,Iancu:2022gpw,Caucal:2024bae}, semi-inclusive DIS \cite{Bergabo:2022zhe,Altinoluk:2024vgg,Caucal:2024vbv}, diffractive structure functions \cite{Beuf:2024msh} and diffractive jets \cite{Iancu:2021rup,Hatta:2022lzj,Iancu:2023lel}, and point to the comprehensive review provided in Ref.~\cite{Morreale:2021pnn}, for further information.

 
\section{Hadronic collisions}\label{sec:HICs}

Hadronic collisions are more complex in nature than the previous examples of DIS. The reason for this is that in hadron-hadron collisions strongly interacting  degrees of freedom are present in both incoming projectiles. Because non-perturbative physics dominates at lower momenta, a general description of hadronic collisions using QCD is not available. For this purpose, we require the use of QCD Effective Field Theories (EFT) such as the CGC. In the context of the latter, particle production in hadronic collisions can be performed in two different kinematic limits, discussed separately in Sec.~\ref{sec:forward} and \ref{sec:dense}. 


\subsection{Forward particle production in p+p/A -- the dilute-dense formalism}
\label{sec:forward}

First, one can choose to study final states (outgoing particles) in the forward rapidity region, where outgoing states are quasi collinear to the direction of the -forward moving- incoming hadron. Because of the kinematic conditions, observables in the forward region are dominated by large-$x$ degrees of freedom in the forward moving hadron, such that collinear factorization is applicable to it. Consequently, particle production in the forward region can be viewed as a process where single partons in the projectile can scatter off the color fields of the backward moving hadron, which is then usually referred to as the target. When the center-of-mass energy is sufficiently high, the target is dominated by small-$x$ gluons requiring a dedicated treatment of multiple parton interactions. Stated differently, this means that in forward kinematics, one is using a single parton coming from a \textit{dilute} projectile to probe the complex small-$x$ wavefunction of a \textit{dense} hadronic target.

Summarizing the above discussion, collinear factorization applies in the dilute projectile, while in the dense target, it is not applicable as interaction with small-$x$ partons need to be taken into account to all orders, just as in the last section. Generally, forward scattering processes can thus be viewed as $p_i + A \rightarrow p_1 + ... + p_n + X $ reactions, where a single perturbative parton $p_i$ from the projectile, exhibits multiple-scattering from the target $A$, producing outgoing partons $p_1 \cdots p_n$, resulting from splittings, radiation off the first incoming parton. 
This semi-perturbative way of computing cross-sections is commonly called the dilute-dense or hybrid formalism in the literature~\cite{Dumitru:2005gt,Albacete:2013tpa,Albacete:2014fwa}.

Clearly, the simplest process one can focus on is single inclusive hadron production in the forward dilute-dense limit, for which the leading order cross-section is given by~\cite{Dumitru:2001jn,Dumitru:2005gt}
\begin{equation}
    \frac{\rmd \sigma^{pA\rightarrow hX}}{\rmd y \rmd ^2\kp_T\rmd^2 \bv }=\int \rmd x\rmd z\frac{1}{z^2} q(x,Q^2_f) \frac{\rmd \sigma^{tot}_{qA}}{\rmd y_q \rmd ^2\qp_T\rmd^2 \bv } D_{q/h} (z,Q_f^2)
    \label{eq:forwardHybrid}
\end{equation}
where we assumed for simplicity that the interacting parton in the projectile is an incoming quark\footnote{Note that a similar expression as in Eq.~\eqref{eq:forwardHybrid} can be obtained for any forward parton interacting with the dense hadronic target. Nevertheless, in the very forward limit, where the cross-section is dominated by large-$x$ partons in the dilute projectile, collinear quarks will dominate over collinear gluons. We will see later that gluons become more relevant in the context of di-jet production.}. The momentum of the produced hadron is related to the momentum of the parton via $\kp_T=z \qp_T$. This formula exemplifies perfectly the three ingredients needed for any forward computation. Starting from the partonic content of the projectile, determined by the non-perturbative collinear quark distribution $q(x,Q_f^2)$, the factor $\sigma^{tot}_{qA}$ represents the cross-section for the interaction of the incoming quark with the target, namely $q_i + A \rightarrow q_f + X $, in our notation above. Eventually, the scattered quark undergoes a fragmentation process to produce the final state hadron, as described by the non-perturbative  fragmentation function $D_{q/h}(z,Q_f^2)$. By following the discussion in Sec.~\ref{sec:CGC}, the cross-section for the interaction of the incoming quark with the target can be computed in the Color Glass Condensate framework as 
\begin{equation}
\frac{\rmd \sigma^{tot}_{qA}}{\rmd y_q \rmd ^2\qp\rmd^2 \bv } = \frac{1}{(2\pi)^2} \tilde{S}(x,\qp, \bv) 
\end{equation}
where $\tilde{S}(x,\qp_T, \bv) = 1-\tilde{D}(x,\qp_T, \bv)$ is the Fourier transform of the dipole operator in the fundamental representation~\cite{Dumitru:2001jn}.

Similarly, to the above example of single inclusive hadron production, the hybrid formalism can be applied to more complex forward observables, such as single and multiple jets~\cite{Dumitru:2005gt,Lappi:2012nh,Chirilli:2012jd,Chirilli:2011km,Lappi:2013zma,Iancu:2020mos} (and references therein), or electromagnetic probes~\cite{Gelis:2002fw,Gelis:2002ki,Benic:2018hvb,Benic:2016uku,Benic:2022ixp}. Generally, the anatomy of the relevant cross-sections remains identical, as the underlying partonic cross-section is calculated within the CGC framework, while the distribution of incoming partons in the projectile is determined by collinear parton distribution functions, just as in perturbative QCD.  When considering hadronic observables, one also needs the knowledge of how the outgoing partons hadronize, which is usually obtained by convolving the cross-section with fragmentation functions $D_{q/h}(z,Q_f^2)$, which are non-perturbative functions extracted  from experiment~\cite{Placakyte:2011az}. Evidently, in the case of multi-parton outgoing states, the convolution with fragmentation functions is needed for each of the outgoing states, as would e.g. be the case for the NLO correction for a quark scattering off a gluonic target, $q+A\rightarrow q+g+X$, where one needs to convolve the cross-section with both $q\to h$ and $g\to h$ fragmentation functions if one is interested in the hadronic end-states.

\subsubsection{\textit{Status Quo}: forward p+A collisions} 
As it was mentioned above, 
the main physical point of choosing final states (outgoing particles) aligned around the forward rapidity region is to enhance the access to the saturated degrees of freedom in the target. 
As the collinear charges move through the cold nuclear matter (CNM\footnote{This is a name used in the literature to call the nuclear effects of the complex nuclear target. It refers to the view of this incoming nucleus as a small, cold "medium" of nuclear matter. The term cold is used in contra-position to the \textit{hot} nuclear medium, the QGP.}) they scatter multiple times from the gluons in the target, modifying the spectra of the charges after the interaction. 

\begin{wrapfigure}{r}{0.5\textwidth}
\includegraphics[width=0.5\textwidth]{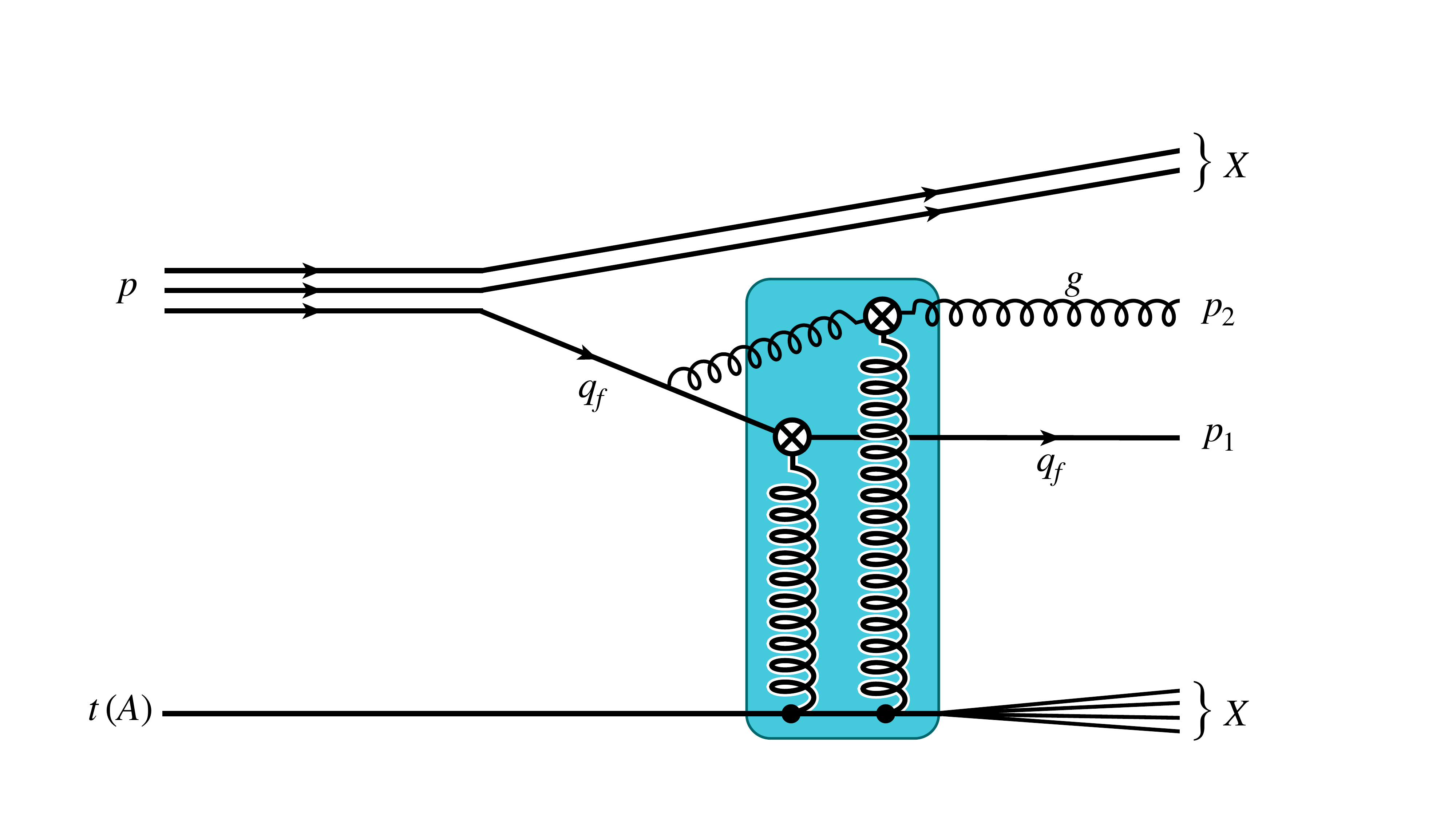}
    \caption{Kinematics of dijet production for the dilute-dense case, in the case of forward $qg$ dijets. Here a quark $q_f$ of the proton emits a gluon $g$ and both interact with the incoming target (t). Through the interaction the quark and the gluon acquire with momenta $p_1$ and $p_2$, respectively.}
    \label{fig:dijetdiagram}
\end{wrapfigure}

 In $p+A$ collisions, the first step is to explore spectra of (un-)identified particles. A useful tool to extract information about the saturation effects of the nucleus in such systems is the so-called nuclear modification factor $R_{\rm pA}$, 
 \begin{equation}
 R_{\rm pA} = \frac{\rmd \sigma_{pA}/\rmd {p_T}\rmd y}{N_{\rm coll} \rmd \sigma_{pp}/\rmd {p_T}\rmd y  }
 \end{equation}
 where particle production in $p+A$ collisions is compared to $pp$, scaled by the number of binary collisions, $N_{\rm coll}$. The $R_{\rm pA}$ has been measured using different nuclear targets (see, e.g. Refs.~\cite{PHENIX:2019gix,ALICE:2012mj}) where a reduction of the total emission ($R_{pA}<1$) was seen for hadrons along the p-going direction.
This is consistent with the physical picture of the CGC, where the multiple-scattering of the forward projectile with the coherent target is expected to cause reduction of particle emission at lower $p_T$~\cite{McLerran:2001sr}, causing a flatter spectrum in comparison to the perturbative $\rmd N/\rmd p_T\rmd y\sim p_T^{-4}$.Much more sensitive signals than single particle spectra are particle correlations~\cite{Lappi:2012nh,Marquet:2007vb,Deganutti:2023qct,Stasto:2011ru,Albacete:2010pg,Iancu:2013dta}. In this case, one can look at the correlations of two forward independent hadrons, $p(p_p)+A(p_A)\to h_1(p_1)+h_2(p_2) + X $, or at the production of two forward jets, $p(p_p)+A(p_A)\to j_1(p_1)+j_2(p_2) + X $, see for example Fig.~\ref{fig:dijetdiagram}. While the measured object itself is not the same\footnote{In the case of hadrons, the partonic cross-section needs to be convoluted with the fragmentation functions of the outgoing partons, just as in eq.~\eqref{eq:forwardHybrid}. For jet production, the parton needs to be evolved by radiating/splitting towards lower momenta, effectively creating a parton \textit{cascade}, or \textit{shower}, which in general will be more sensitive to final state effects.}, both present the same initial mechanism,  
namely the particle production through radiation or splitting from the incoming forward parton.

\begin{figure}
    \centering
 \includegraphics[width=0.4\linewidth]{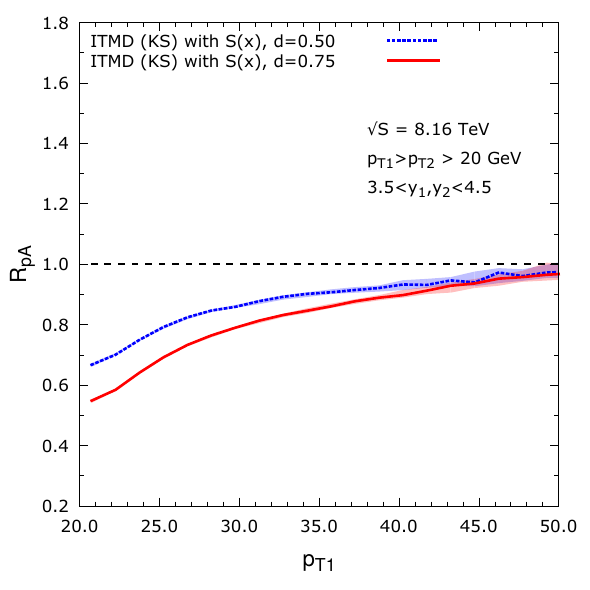} \includegraphics[width=0.4\linewidth]{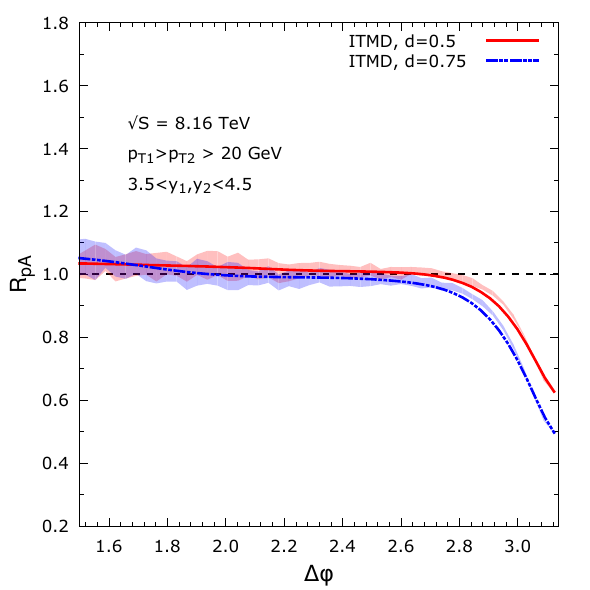}
    \caption{ (\textit{Left}) Nuclear modification factors as a function of the transverse momentum of the leading jet.(\textit{Right}) Nuclear modification factors of the angular distribution for two jets, shown for two values of the nuclear saturation scale. At large angles, the p+A dijet angular distribution sees an effective suppression. Here, the parameter $d$ estimates the uncertainty in the strength of saturation in Pb nuclei (as deviation from $Q_A^2 \sim A^{2/3} Q_p^2$). Figures taken from Ref.~\cite{vanHameren:2016ftb}.}
    \label{fig:dijet}
\end{figure}

The initial parton restricts, through the initial hard scatterings/splittings, the nature of the initial partons for the jets ($j_1$ and $j_2$). On the other hand, multiple scatterings with the soft gluons within the target determine the transverse momentum imbalance between the two jets, $\boldsymbol{q}_T=\boldsymbol{p}_{1,T}- \boldsymbol{p}_{2,T}$. The crux of this point is that, as $|q_T|$ approaches the saturation scale of the target, the dijet correlation becomes sensitive to saturation effects in the gluon distributions. Intuitively, the partonic pair has the  "relative size" large enough to multiply interact with a highly occupied set of gluons. The first effect of this is that the transverse momentum broadening caused by such interactions causes the suppression of the correlation at small relative momenta. Additionally, the deflection of the pair causes a broadening of the angular distribution of the two particles, where the back-to-back peak (at $\Delta \phi=\pi$) becomes softer and broadened.
In Fig.~\ref{fig:dijet}, taken from Ref.~\cite{vanHameren:2016ftb}, the reader can find the saturation effects as visible in the nuclear modification factor for forward dijet production, as computed in the improved TMD factorization formalism~\cite{Kotko:2015ura}, which, loosely speaking, corresponds to the $|\boldsymbol{p}_{1/2,T}|\gg Q_s $ limit of the CGC EFT \cite{Fujii:2020bkl,vanHameren:2023oiq}. In the left plot, the reader can see an overall suppression of the jet production, which becomes stronger with decreasing transverse momentum $p_{T}^{1}$ of the leading jet. On the right side, we also have a nuclear modification factor, but now for angular distributions. One can read from this image the aforementioned effect of the suppression of back-to-back correlations. This effect, predicted in Ref.~\cite{Marquet:2007vb}, was later experimentally found at RHIC  by the PHENIX collaboration~\cite{PHENIX:2011puq,Braidot:2010zh}. Additionally, LHC has measured dijet yields at forward rapidities in proton-proton and proton-lead collisions,
although the search for saturation signals in these observables was not conclusive enough~\cite{ATLAS:2019jgo}.

A particularly powerful tool to explore the gluon wave-functions of the initial state are electromagnetic probes, photons and dileptons. Devoid of strong interactions, the EM probes created by interactions with the saturated target are themselves the final states observed. 
Measurements of forward isolated photons have been collected by
the ATLAS and CMS collaborations for moderate rapidities in proton-proton collisions and for
intermediate transverse momentum~\cite{ATLAS:2010uco,CMS:2012oiv,CMS:2010svd,CMS:2011nkw}. Photon (dilepton) production has been computed for dilute dense system at LO in Ref.~\cite{Gelis:2002ki,Gelis:2002fw}. The extension to NLO -in the CGC power counting scheme- has been performed in Refs.~\cite{Benic:2016uku,Benic:2018hvb,Benic:2016yqt,Jalilian-Marian:2012wwi}\footnote{While NLO parton emission has been computed in Refs.\cite{Roy:2019hwr,Roy:2019cux} for the equivalent mechanism in DIS, NLO emission is still not fully understood in the dilute-dense limit of the CGC. However, NLO corrections to LO channel have been performed in the TMD factorization limit by including gluon emission\cite{Altinoluk:2018byz}}. Because of the kinematics of gluon fusion, the NLO is more relevant at lower rapidities, the overlap regions at intermediate rapidities are an interesting case to explore the sensitivity to the  gluon distributions in the projectile. Just as in the dihadron case, photon-hadron correlations are more sensitive to saturation physics~\cite{Benic:2022ixp}.

\subsubsection{New developments \& open questions}
\label{sec:dil-dense-new}

While the most promising phenomenological signatures of small-$x$ gluon saturation remain the same, over the last decade, the field has worked towards achieving higher-order precision, which involves developing concurrent corrections both to the individual cross-sections and to the $x$-evolution of the gluon distributions. A significant amount of progress has been achieved to describe hadronic collisions in the dilute-dilute and dilute dense to NLO accuracy  on the side of the impact factors, both on single particle production formulas \cite{Chirilli:2011km,Chirilli:2012jd,Altinoluk:2014eka,Ducloue:2016shw} and for dijet production in pA~\cite{Iancu:2020mos}. Early works famously saw negative cross-sections at large-$p_T$ \cite{Stasto:2013cha,Altinoluk:2014eka,Ducloue:2016shw}, a problem which has received two independent solutions~\cite{Shi:2021hwx,Iancu:2016vyg}. The reader can find a phenomenological analysis using the solution \cite{Iancu:2016vyg} in Ref. \cite{Ducloue:2017mpb}.
On the other hand, quite some effort has gone to improve the small-$x$ evolution equations from the well established leading logarithm (LL) level --including running coupling effects -- to the NLL accuracy, effects which have been derived for the BK~\cite{Balitsky:2007feb} and JIMWLK equations~\cite{Iancu:2020whu,Kovner:2013ona,Balitsky:2013fea,Lublinsky:2016meo,Kovner:2014lca}. Additionally, NLL corrections to the JIMWLK evolution have been recently  derived also for massive quark corrections in Ref.~\cite{Dai:2022imf}.
It is important to note also that the numerical implementation of these equations has proven to be a complex task, where e.g. numerical instabilities on the solution of NLL BK equations were found for phenomenologically relevant initial conditions, and later fixed by additional resummations~\cite{Beuf:2014uia,Iancu:2015vea,Ducloue:2019ezk}. Obtaining full numerical evolution of the JIMWLK equation at NLL is still an active area of research \cite{Lappi:2016fmu}.

So far a sophisticated modeling of nuclear geometry and multiplicity dependent effects has not been performed for many dilute-dense CGC calculations. However, it has recently been shown in Ref. \cite{Deganutti:2023qct}, that nuclear structure modeling in the gluon distributions leadseems to a dependenchave an effect on the saturation scale different fromly thean naively expected from $Q_s^2\sim A^{1/3}$, altering the naive expectations (and computations) based on such behavior. Since forward particle production provides a natural way of constraining our knowledge of the initial small-$x$ many-body wavefunction of the colliding hadrons, improvements in this respect would be also desirable in order to reduce theoretical uncertainties in the initial deposition
of charge and energy in heavy-ion collisions (see Sec.~\ref{sec:dense}) and improve the predictive power of the entire framework. 


\subsection{Initial state of Heavy-Ion Collisions (HICs) -- the dense-dense formalism}
\label{sec:dense}
High-energy heavy-ion collisions produce a complex QCD many-body system, whose space-time evolution is conventionally described within multi-stage evolution models~\cite{Elfner:2022iae}, including the energy deposition and early-time dynamics in the initial state, the viscous hydrodynamic evolution of the Quark-Gluon Plasma (QGP), the re-hadronization of the QGP as well as the subsequent re-scattering and decay of produced hadrons. In this context the CGC framework provides a QCD approach to compute the properties of the initial state, from the underlying structure of hadrons and nuclei. Stated differently, due to the complex reaction dynamics, the scope of CGC calculations is not to describe the final state particle spectra observed in heavy-ion collisions, but merely to obtain the space-time structure of the initial state shortly after the collision, which in turn can then be used as a starting point for describing the subsequent reaction dynamics.

In high-energy collisions kinematics dictates that the energy deposition around mid-rapidity is dominated by small-$x$ gluons, which are highly abundant in the colliding nuclei.  Both nuclei are then typically considered as "dense" color charged objects, and described in terms of counter-propagating eikonal color currents~\cite{McLerran:1993ka}
\begin{equation}
 J_{A/B}^\mu = \delta^{\mu\pm}\delta(x^\mp)\,t^a\rho^a_{A/B}(\xv_T)\,.
    \label{eq:currents}
\end{equation}
localized around the $x^{\pm}=0$ axis, where the subscript $A/B$ refers to the two incoming nuclei moving along the positive/negative $\pm z$ direction. In order to describe the ensuing dynamics we will use Milne coordinates, 
\begin{equation}
    x^{\pm} = \frac{\tau}{\sqrt{2}}e^{\pm\eta_s} \quad\Rightarrow\quad \tau = \sqrt{2\xp x^-}\,,\,\eta_s = \frac{1}{2}\log\left[\frac{x^+}{x^-}\right]\,,
\end{equation}
where $\tau$ is the longitudinal proper time and $\eta_s$ is the space-time rapidity. We also note in passing that in terms of the Milne coordinates, the fields are now given by 
\begin{equation}
\tau\,A^{\tau} = x^+A^-+ x^-A^+ \quad \text{and}\quad \tau^2 A^{\eta_s} = x^+A^- -x^-A^+ 
\end{equation}
while the transverse components $A^{i}$ are unaffected by the transformation.

Developing a first principles description of the real-time non-equilibrium dynamics of QCD is an outstanding problem~\cite{Berges:2020fwq}. However, at leading order in the strong coupling constant the ensuing dynamics of the collision are described by the solution to the classical Yang-Mills (CYM) equations~\cite{Gelis:2007kn}

\begin{equation}
    D_\mu F^{\mu\nu}=J^\nu 
    \label{eq:YM}
\end{equation}
where in Fock-Schwinger $(A^{\tau}=0)$ gauge, the eikonal color currents are individually conserved $(D_{\mu}J^{\mu}_{A/B}=0)$, such that  $J^\mu = J_A^\mu + J_B^\mu$ denotes the combined color current of both colliding nuclei\footnote{If this is not the case, the covariant conservation equation $D_{\mu}J^{\mu}=0$ needs to be solved concurrently with the class. Yang-Mills equations, with boundary conditions describing the incoming nuclei prior to the collision~(see e.g.~\cite{Ipp:2021lwz,Ipp:2024ykh,Ipp:2020igo,Ipp:2021drn,Schlichting:2020wrv}).}. Next-to-leading order corrections include quantum fluctuations on top of the CYM equations~\cite{Schenke:2012wb,Dusling:2011rz,Epelbaum:2013waa,Gelis:2016upa}, however a consistent NLO description is yet to be developed.

\begin{wrapfigure}{l}{0.35\textwidth}
\includegraphics[width=0.35\textwidth]{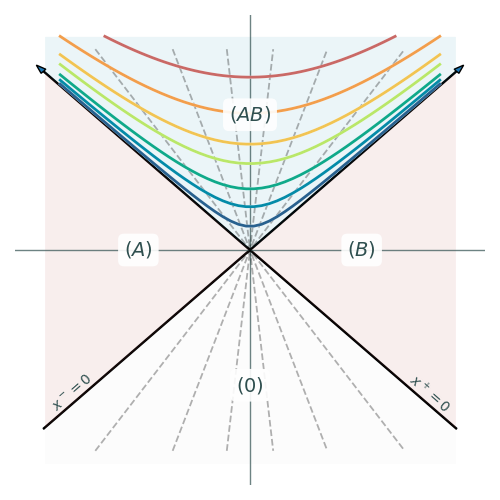}
    \caption{Space-time diagram for the evolution of a high-energy heavy-ion collisions \label{fig:minkowskidiagram}}
\end{wrapfigure}

The solution to Eq.~\eqref{eq:YM} can be conveniently expressed by subdividing space-time in four regions 
\begin{equation}
\begin{split}
A^\mu(x) =& \theta(-x^-)\theta(-x^+)\, A^\mu_0(x)+\theta(x^-)\theta(-x^+)\, A^\mu_{\rm A}(x) \\
& + \theta(-x^-)\theta(x^+) A^\mu_{\rm B}(x)+\theta(x^-)\theta(x^+) A^\mu_{\rm AB}(x)
\end{split}
\label{eq:subregions}
\end{equation}
depending on whether or not causal contact with nucleus $A$ and $B$ has been established, as illustrated in Fig.~\ref{fig:minkowskidiagram}. 
Starting from the classical vacuum $A^{\mu}_{0}(x)=0$ before the collision, the solutions in regions $A$ and $B$ are the same as in the case of a single nucleus, such that the field strength $F^{\mp i}$ is located inside the nuclei at $x^{\pm}\simeq 0$, and the non-vanishing components of the gauge field in Fock-Schwinger gauge are given by
\begin{equation}
    A^{i}_{A}(x)=\frac{i}{g} V_{A}^{\dagger}(x) \partial^{i}V_{A}(x)\quad {\rm and} \quad  A^i_{B}(x)=\frac{i}{g} V_{B}^{\dagger}(x) \partial^{i}V_{B}(x)\,.
\end{equation}
where $V(x)$ denote the light-like Wilson lines as in Eq.~\eqref{eq:wilson_line}.\footnote{Evidently, the solution in Fock-Schwinger (FS) gauge can be directly obtained from the individual solutions~\eqref{eq:single_nucleus_cgc} in light-cone (LC) gauge, by performing
a gauge transformation $A^{\mu}_{FS}(x)=G^{\dagger}(x)A^{\mu}(x)G(x)+\frac{i}{g} G^{\dagger}(x) \partial^{\mu}G(x)$ with $G(x)= \mathcal{P} \exp ( \int_{-\infty}^{x^{+}} dz^{+} ig A^{-}(\xp,\xt)) \mathcal{P} \exp ( \int_{-\infty}^{x^{-}} dz^{-} ig A^{+}(\xm,\xt))$.} Due to the non-linear nature of the CYM equations, analytical solutions are no longer available for the evolution in the forward light-cone(AB). 
Strikingly, however, the initial conditions for the evolution in the forward light-cone (AB), at $\tau=0^{+}$, immediately after the collision, can still be determined in a closed form~\cite{Kovner:1995ja,Kovner:1995ts}, 
\begin{equation}
\begin{split}
        A^{\eta_s}(\tau\rightarrow 0^+,\eta^{s},\xv_T ) &= \frac{\rmi g}{2} \left[A^{i}_A(\xv_T),A^{i}_B(\xv_T)\right]\\  A^{i}(\tau\rightarrow 0^+,\eta^{s},\xv_T ) &= A^{i}_A(\xv_T)+A^{i}_B(\xv_T)
\end{split}
\label{eq:Glasma0}
\end{equation}
which represent boost-invariant longitudinal chromo-electric and chromo-magnetic fields, mediating the color exchange between the colliding nuclei~\cite{Lappi:2006fp}. Since the $\tau$-derivatives of the fields can be found to vanish at $\tau\rightarrow 0$, the source-free classical Yang-Mills (CYM) equations~\eqref{eq:YM}, along with the gauge fields in Eq.~\eqref{eq:Glasma0} represent an initial value problem for the space-time evolution in region (AB).

Starting from the initial ``Glasma'' fields in Eq.~\eqref{eq:Glasma0}, different aspects of the subsequent space-time dynamics can be studied (semi-) analytically based on perturbative~\cite{Lappi:2017skr} or proper time expansions~\cite{Carrington:2021qvi,Carrington:2020ssh}, or numerically using real-time lattice gauge theory techniques to solve the CYM equations in the forward light-cone~\cite{Krasnitz:1998ns,Lappi:2003bi,Lappi:2011ju}.By consistently following the perturbative approach, one recovers factorization formulas for inclusive particle production in the dilute limit~\cite{Dumitru:2001ux,Lappi:2017skr}, which provide the basis for the {\Dipper} initial state model of heavy-ion collisions \cite{Garcia-Montero:2023gex}. Numerical simulations of the CYM equations underlie the phenomenologically successful IP-Glasma model~\cite{Schenke:2012hg,Schenke:2012wb}, which employs the color charge configurations of the IP-Sat model described in Sec.~\ref{sec:CGC}, to compute the energy-momentum tensor of the Glasma, 
\begin{equation}
	T^{\mu\nu} = -F^{\mu\rho}\,F^{\nu}_\rho+ \frac{1}{2} g^{\mu\nu} F_{\lambda\sigma}F^{\lambda\sigma} \,.  
\end{equation} 
Initially the longitudinal chromo-electric and chromo-magnetic Glasma flux tubes give rise to the following diagonal form of the energy-momentum tensor
\begin{equation}
\left. T^{\mu}_{~\nu}(\tau=0^{+}) \right|_{\text{Glasma}}= \text{diag}(e_0,e_0,e_0,-e_0)
\end{equation} 
which describes a non-equilibrium system in its rest frame, with local energy density $e_{0}(x)=T^{\tau}_{~\tau}(x)$ and negative longitudinal pressure $\tau^2T^{\eta\eta}=-e_{0}(x)$. However, on a time scale $\tau \sim 1/Q_s$ phase decoherence of the classical fields rapidly leads to a different form of the energy-momentum tensor
\begin{equation}
\left. T^{\mu}_{~\nu}(\tau_{\text{EKT }}=1/Q_{s}) \right|_{\text{EKT }} \simeq \text{diag}(e,e/2,e/2,0)\label{eq:EST_after_classical}
\end{equation}
which describes a system of weakly interacting quasi-particles, with transverse momenta $\sim Q_s$ and vanishing longitudinal momenta in the local rest frame $u^{\mu}$, which neglecting the off-diagonal components of $T^{\mu\nu}$, related to transverse expansion (pre-flow), is given by $u^{\mu}=(1,0,0,0)$ in Milne coordinates. Despite the fact that at this point the system is still highly anisotropic, the initial state in Eq.~\eqref{eq:EST_after_classical} is successfully used as initial conditions for the subsequent hydrodynamic evolution of the QGP~\cite{Gale:2012rq}. In this case, the system is still far from equilibrium at the beginning of the hydrodynamics phase, and the evolution towards equilibrium is then described macroscopically by \emph{viscous} hydrodynamics. Since at $\tau \sim 1/Q_s$, the system can be described by weakly interacting quasi-particles, one alternative is to describe its evolution towards equilibrium using QCD kinetic theory~\cite{Baier:2000sb,
		Kurkela:2015qoa,Keegan:2015avk}, which naturally leads to a near-equilibrium form of the energy momentum tensor
\begin{equation}
 \left. T^{\mu}_{~\nu}(\tau=\tau_{\rm th}) \right|_{\text{Hydro}}= \text{diag}(e,e/3,e/3,e/3)\,+ \text{viscous corrections}.
 \end{equation}
and also describes the evolution of pre-equilibrium flow~. Evidently, understanding the microscopic mechanisms behind equilibrium and the onset of hydrodynamic behavior is one of the most outstanding challenges in contemporary heavy-ion physics, and we refer to~\cite{Berges:2020fwq,Schlichting:2019abc} for dedicated reviews on this subject. Nevertheless, it is important to properly describe the pre-equilibrium stage, as the duration of the pre-equilibrium stage determines the overall amount of energy deposition and also has a non-trivial impact on the collision geometry~\cite{Giacalone:2019ldn}.
Since the CGC initial state models are based on microscopic QFT calculations, they also give access to momentum space information and the possibility to include initial state momentum correlations~\cite{Giacalone:2020byk}. However, such correlations will then be destroyed by subsequent space-time evolution of the QGP~\cite{Greif:2017bnr,Schenke:2019pmk}, they are relatively short range in rapidity~\cite{Schenke:2022mjv} and decrease with increasing system size~\cite{Mace:2018vwq,Schenke:2015aqa} and may thus only be relevant in short-lived small collision systems.

Generally, initial state models based on saturation physics provide not only the spatial geometry, but also the overall magnitude and center-of-mass energy dependence of energy deposition in the initial state of heavy-ion collisions.
Beyond the concrete implementations of these ideas in the IP-Glasma~\cite{Schenke:2012wb}\footnote{Based on the IP-Sat model and real-time lattice gauge theory simulations of the CYM equations.}, the {\Dipper}~\cite{Garcia-Montero:2023gex}\footnote{Based on the IP-Sat model and a $k_T$-factorized approximation to the CYM equations.}, as well as the saturation inspired EKRT model~\cite{Niemi:2015qia,Kuha:2024kmq}\footnote{Based on collinear factorization with final state saturation imposed.}, the insights from such microscopic calculations have also proven valuable for the development of empirical initial state models. Such is the case of TrENTo~\cite{Moreland:2014oya}, where energy and charge deposition in heavy-ion collisions are determined according to relatively simple parametric formulas. Notably, a detailed statistical analysis of heavy-ion data seems to support the general dependence inferred from saturation models~\cite{Nijs:2020ors}. Since the CGC models are independently constrained from DIS experiments, the initial state of heavy-ion collisions is actually reasonably well constrained, meaning that comparisons to experimental heavy-ion data should provide rather stringent constraints on the space-time evolution of the QGP, that connects the initial state to experimetnal measurements. However in practice this potential has not been fully exploited, as it is common practice in heavy-ion phenomenology to re-adjust e.g. the normalization of the initial state energy, in order to reproduce measured particle spectra.

It is also important to note that since the models are based on QCD structure of nucleons, they are not limited to heavy-ion collisions but in principle applicable to collisions of all hadronic species at sufficiently high center of mass energy. In particular for small systems, effects of sub-nucleonic structure resulting in a non-trivial proton geometry, are in fact essential for the description of collective flow in p+A collision~\cite{Bozek:2017fdv}, albeit it is not clear to what extent a hydrodynamic description of the space-time evolution is quantitatively accurate in small systems~\cite{Ambrus:2022qya}. Conversely, in large collision systems, such as A+A collisions - with Au, Pb nuclei -- the  relevant aspects of geometry are determined by fluctuations at the largest scales, namely the nucleon positions. Since in this case, the most relevant fluctuations are set at distances of 1-10~fm, correlations of low-energy  nuclear structure start becoming important, and there are ongoign efforts to explore effects of $\alpha$-clustering~\cite{Cao:2023kbs,Ding:2023ibq}, ground state deformation~\cite{Bally:2023dxi,Giacalone:2024ixe,Zhang:2024vkh} and the appearance of a neutron skin~\cite{Giacalone:2023cet,Pihan:2024lxw} for different nuclear species.


\begin{figure}
    \centering
\includegraphics[scale=0.20]{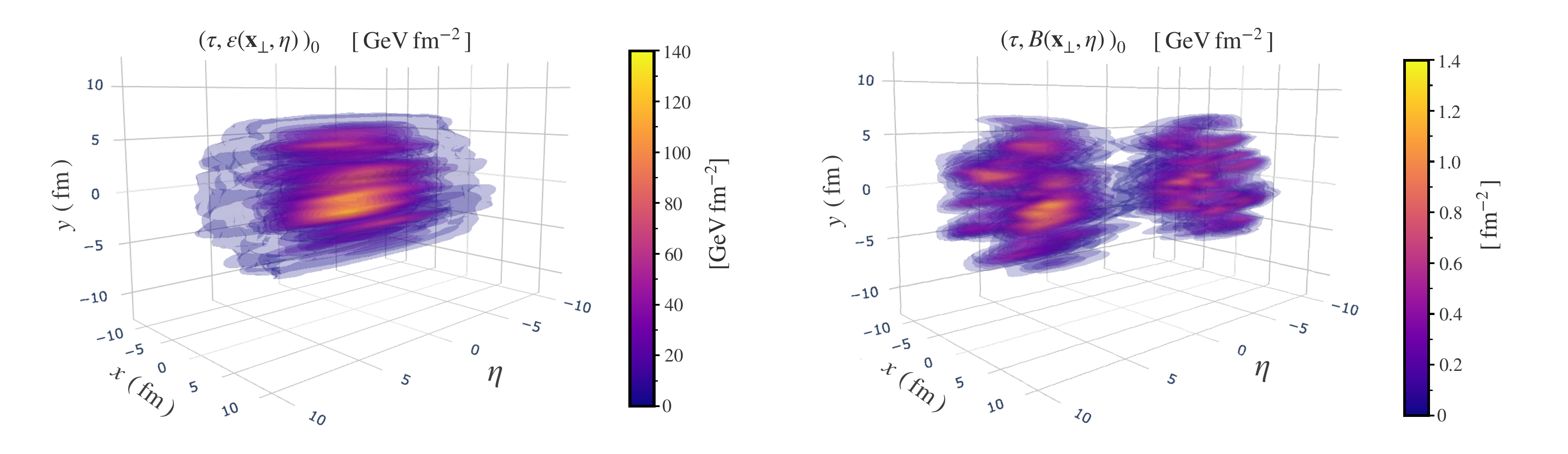}
    \caption{Deposition of energy density (\textit{left}) and baryon charge (\textit{right}) for a single 3D event. The charge and energy deposition is computed in the initial state \Dipper{} framework. Figure taken from Ref.~\cite{Garcia-Montero:2023opu}.}
    \label{fig:Promoplot}
\end{figure}

\subsubsection{New developments \& open questions}
\label{sec:HICsOpen}
Currently one the most important challenges in describing the initial state of heavy-ion collisions, is the development of a first principles understanding of the $3+1D$ space-time structure of the initial state. Generally, the CGC provides a natural framework to address this question, however so far it has proven challenging to develop a consistent formulation of the 3+1D dense-dense collisions.

By extending the support of the eikonal color currents in Eq.~\eqref{eq:currents} to a finite longitudinal thickness, different implementations of a $3+1$ D coordinate space picture of the Glasma have been put forward~\cite{Gelfand:2016yho,Ipp:2017lho,Schlichting:2020wrv,Ipp:2021lwz,Ipp:2024ykh,Matsuda:2024mmr,Matsuda:2024moa}. However, it is not yet clear how to fully relate the longitudinal structure of the eikonal color currents to the small $x$ structure of the colliding nuclei. While this relation should partially come from the JIMWLK evolution, other aspects, e.g. the sources finite size\cite{Schlichting:2020wrv,Ipp:2021lwz}, may have an important impact.
Additionally, it remains to be seen how genuine sub-eikonal corrections~\cite{Altinoluk:2015xuy,Altinoluk:2024dba} could be incorporated into this framework.

Phenomenological saturation models for the 3+1D initial state of heavy-ion collisions, instead employ a momentum space picture, which emerges in the dilute-dense limit, where it becomes possible to establish a relationship between the momentum fractions $x$ of incoming partons and the momentum rapidity $y$ of produced partons~\cite{Kharzeev:2001yq,Kharzeev:2002ei,Gelis:2003vh,Garcia-Montero:2023gex}. One example is the {\Dipper} model~\cite{Garcia-Montero:2023gex,Garcia-Montero:2025bpn}, which is phrased in the language of TMDs, where e.g. the initial charge deposition due to valence quark stopping is determined by the probability of the valence charge acquiring transverse momentum from  the interaction with saturated gluons in the target (see \cite{Garcia-Montero:2024jev})\footnote{By expanding in the transverse momentum transfer from the target to the projectile is, one can recover the collinear factorization limit, in which models such as EKRT also provide a 3D picture of the initial state~\cite{Kuha:2024kmq}.}.

Notably in the momentum space picture, it is in principle straightforward to include the energy and charge deposition due to quarks, as done e.g. in the {\Dipper}, as well as higher order perturbative corrections, albeit this has not been accomplished to date. By identifying space-time and momentum rapidity one then obtains a coordinate space picture of the 3D initial state, as illustrated in Fig.~\ref{fig:Promoplot}.

We finally note that beyond the development of concrete implementations, there are more fundamental questions regarding the validity of factorization assumptions when considering un-equal rapidity correlations \cite{Iancu:2013uva}, which will require further attention in the future.

\section{Concluding Remarks}\label{sec:conclusions}

Effective theories for high-energy nuclear interactions, provide important insights into the behavior of nuclear matter under extreme conditions. 
By modeling the complex gluon dynamics at small-$x$, the Color Glass Condensate (CGC) framework represents a significant step towards a unified description of high-energy scattering processes in QCD, bridging experimental observations in lepton-hadron, hadron-hadron and hadron-nucleus collisions with theoretical models. While leading order cross-section calculations in the CGC provide an intuitive understanding of the underlying physics picture, 
phenomenological studies based on higher order calculations will be key in refining our understanding of QCD at high CoM energies and enhancing predictive capabilities for both current and next-generation colliders. Beyond the precision frontier, new developments, especially those addressing longitudinal correlations and the three-dimensional many-body structure of hadrons, will be instrumental for the development of a cohesive physical picture of nuclei at high energies. Even though it is presently unclear when or to what extent a smoking gun signal for gluon saturation at small $x$ will be discovered at present or future facilities, it is clear that the general theoretical framework of saturation physics will continue to provide important contributions to unraveling the fundamental structure of nuclear matter at extreme scales.

\bmhead{Acknowledgments}
We thank B.~Schenke and the referees of EPJA for execellent comments and suggestions.
The authors acknowledge support by the Deutsche Forschungsgemeinschaft (DFG, German Research Foundation) through the CRC-TR 211 ‘Strong-interaction matter under extreme conditions’-project number 315477589 – TRR 211. 






\bibliography{References}

\end{document}